# No evidence of "isostructural electronic transitions" in compressed hydrogen

L. Dubrovinsky[a], N. Dubrovinskaia[b*] & M. I. Katsnelson[c]

[a]Bavarian Research Institute of Experimental Geochemistry and Geophysics, University of Bayreuth, D-95440 Bayreuth, Germany;

[b]Material Physics and Technology at Extreme Conditions, Laboratory of Crystallography, University of Bayreuth, D-95440 Bayreuth, Germany;

[*]email: natalia.dubrovinskaia@uni-bayreuth.de

[c]Institute for Molecules and Materials, Radboud University, Heyendaalseweg 135, 6525AJ Nijmegen, The Netherlands

**Ji et al.[1] report performing X-ray diffraction on hydrogen compressed to over 250 GPa. It is a remarkable technical achievement. However, the experimental data presented and discussed in the paper do not support the main conclusion that hydrogen undergoes an isostructural phase transition and preserves the hexagonal close packed (*hcp*) structure up to the highest pressure achieved. The behavior of compressed hydrogen in the studied pressure range cannot be explained by electronic topological transition (ETT), as claimed in the paper.**

Close inspection of the presented XRD images (they are two, in Fig. 1b and Fig. 2b) collected at pressures of 212 GPa (phase III) and 232 GPa (phase IV), as defined by the authors, shows numerous unidentified diffraction spots (marked by yellow rectangles in our **Fig. 1 below**). The sizes, shapes and appearance of these unidentified spots are similar to the spots assigned by the authors to hydrogen, whereas they are remarkably different from reflections, which could tentatively, based on our own experience, be assigned to reflections from the diamond anvils (see **Fig. 1 below**). We are fully convinced that a conclusion, whether hydrogen remains *hcp* or undergoes a structural transformation, could be drawn only if all spots observed by Ji et al. in their plentiful XRD patterns were explained one-by-one (that is, however, not the case). An example of the careful spot-by-spot analysis of a rich single-crystal diffraction pattern can be found in Ref. 2- see Fig. 2 therein).

Ji et al.[1] claim that they performed a single-crystal X-ray diffraction (SXRD) study of solid hydrogen, whereas in fact they did not. They merely only determined the *d*-spacings from the positions of individual diffraction spots. The SXRD includes the data collection in a way, which allows the reconstruction of the reciprocal space, and the data analysis, which allows identifying diffraction peaks produced by a unique single crystal (or by a single-crystalline grain in a polycrystalline sample), indexing the peaks (i.e. determination of the orientation matrix and the lattice parameters), finding the symmetry of the crystal, extracting the intensity of the peaks, and eventually solving and refining the crystal structure. It may happen that the experimental environment or the quality of material do not allow to realize the final steps of the SXRD analysis (i.e. to solve and/or refine its crystal structure), but all previous steps must be followed in order the study could be called the single-crystal XRD. If Ji et al.[1] would have done SXRD and would have shown reconstructed reciprocal planes confirming the absence of unidentified reflections, then there wouldn't be a question regarding the interpretation of their diffraction pattern. As scientific terminology carries the information about the type of investigation performed, which implies a possibility to judge the reliability and scientific value of the results, the question "was the SXRD performed" is principal. Here the answer is a definitive "no".

An additional issue, which puts under scrutiny the conclusion regarding the phase transition, is the internal inconsistency of the procedure used to calibrate pressure and to detect pressure-induced structural changes. Determination of pressure is always a concern in the multimegabar pressure range; the authors write: "the calibrated pressure-dependent shift of the $d_{100}$ value of solid $H_2$ was used as the pressure scale". In the **Extended Data Fig. 5** they show the calibration curve ($d_{100}$ *vs* P), which is validated up to 160 GPa due to the Au pressure marker used. Then, in the experiments at pressures above 160 GPa, *in the absence of Au*, the $d_{100}$ value of solid $H_2$ is continued to be used for pressure determination in order to plot the pressure dependence of $d_{100}$, $d_{002}$, and $d_{101}$ in **Fig. 2e** (at numerous pressure points the reflection (002) is missing). It is not a surprise that both the $d_{100}$ (**Fig. 2e**) and the unit cell parameter *a* (**Fig. 3a**) of solid hydrogen show a smooth monotonous variation with pressure above 160 GPa, as for the hexagonal structure the unit cell parameter *a* is directly expressed through $d_{100}$ ($1/d_{100}^2 = 4/3a^2$). This also implies that, in fact, the pressure is first calibrated as a function of the unit cell parameter *a* of solid hydrogen, but then the ratio *c/a*, which includes the same parameter *a*, is used to conclude about pressure-induced structural changes. One cannot use the same parameter simultaneously for the pressure calibration and for the detection of pressure-induced structural changes. In other words, an extrapolation of the calibration curve $d_{100}$ *vs* P above 160 GPa was made under the assumption that above 160 GPa there are no structural transformations.

There is also a problem with the agreement between pressures determined from "hydrogen" and "diamond Raman edge" scales. First, it is unclear why "diamond Raman edge" pressure values are not provided for all data points, at least for comparison. Second, although the difference seems to be small (only 2 GPa, according to **Extended Data Fig. 4**), it raises a question, whether the tiny effect, attributed by authors to be the evidence of the isostructural phase transition, could be observed at all. Indeed, if the "hydrogen" scale gives "232 GPa" and on the "diamond Raman edge" scale it is "234 GPa" (see **Extended Data Fig. 4**), but, in fact, in **Fig. 2e** this point is shown at about 228 GPa (see **Fig. 2e**), then the claim that an irregularity is observed in the compressional behavior of hydrogen above this pressure seems to be a consequence of a problem with an objective characterization of pressure and the graphical presentation of the data.

Ji et al.[1] suggest that the isostructural transition and anomalous behavior of *c/a* ratio above 230 GPa is a consequence of an electronic topological transition (ETT). According to all available experimental information, at ~240 GPa and room temperature hydrogen is not metallic, and according to Ji's et al.[1] own calculation, there are no electrons on Fermi level in phases III and IV in the same pressure range. At the same time, the existing theory of anomalies of lattice properties associated to ETT[3] assumes without any doubt that it is the coincidence of Van Hove singularity with the Fermi energy that results in the anomalies. Thus, a discussion of the ETT in relation to hydrogen behavior in a semiconducting state is misleading.

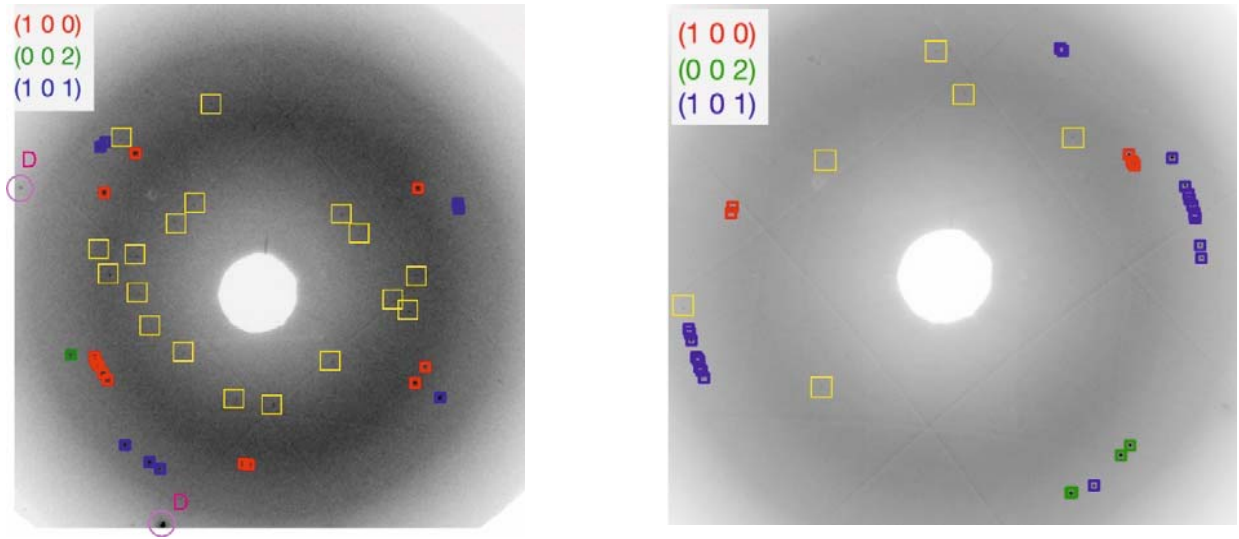

**Fig. 1.** Copies from Ji et al.[1] of "merged raw XRD images showing XRD spots" collected at 212 GPa (left - original Fig. 1b) and 232 GPa (right- original Fig. 2b). Yellow squares mark spots not explained by Ji et al.[1]. Pink circles, designated also with the letter "D", mark the spots which we think to be due to diamond of the anvils.

**References**

1. Ji, C. et al. Ultrahigh-pressure isostructural electronic transitions in hydrogen. *Nature* **573**, 558-562 (2019).
2. Ismailova, L. et al. Stability of Fe, Al-bearing bridgmanite in the lower mantle and synthesis of pure Fe-bridgmanite. *Sci. Adv.* **2**, e1600427 (2016).
3. Katsnelson, M. I., Naumov, I. I. & Trefilov, A. V. Singularities of the electronic structure and premartensitic anomalies of lattice properties in beta-phases of metals and alloys. *Phase Transitions* **49**, 143-191 (1994).